\documentclass[runningheads]{llncs}                        

\usepackage{url}
\usepackage{makecell}
\usepackage{microtype}                 
\PassOptionsToPackage{warn}{textcomp}  
\usepackage{textcomp}                  
\usepackage{mathptmx}                  
\usepackage{times}                     
\usepackage{cite}                      
\usepackage{tabu}                      
\usepackage{booktabs}                  
\usepackage[acronym]{glossaries}
\usepackage{graphicx}
\usepackage{tikz}
\usetikzlibrary{arrows.meta, positioning}

\newacronym{cae}{CAE}{Computer-Aided Engineering}
\newacronym{ic}{IC}{Inclusion Criteria}
\newacronym{ec}{EC}{Exclusion Criteria}
\newacronym{ai}{AI}{Artificial Intelligence}

\graphicspath{{figures/}{pictures/}{images/}{./}} 
 
\usepackage{xcolor}
\usepackage{amssymb,amsfonts}
\makeatletter
\newcommand{\ygg@basicalert}[2]{\fbox{\bfseries\sffamily\scriptsize#1}{\color{red}\sf\small$\blacktriangleright$\textit{#2}$\blacktriangleleft$}}
\newcommand{\LAYAN}[1]{\ygg@basicalert{LAYAN}{#1}}

\begin{document}

\title{AI for Better UX in Computer-Aided Engineering: Is Academia Catching Up with Industry Demands? A Multivocal Literature Review}

\author{Choro Ulan uulu\inst{1}\textsuperscript{*}\orcidID{0009-0009-3748-3273} \and
Mikhail Kulyabin\inst{1}\orcidID{0009-0007-0440-030X} \and \\
Layan Etaiwi\inst{1}\orcidID{0000-0001-9250-7578} \and
Nuno Miguel Martins Pacheco\inst{1}\orcidID{0000-0002-2198-7823} \and
Jan Joosten\inst{1}\orcidID{0009-0003-2503-4741} \and
Kerstin Röse\inst{1}\ \and
Filippos Petridis\inst{1}\ \and
Jan Bosch\inst{2,3}\orcidID{0000-0003-2854-722X} \and \\
Helena Holmström Olsson\inst{4}\orcidID{0000-0002-7700-1816}}
\authorrunning{C. Ulan uulu et al.}

\titlerunning{AI for Better UX in Computer-Aided Engineering}

%
\institute{Siemens AG, Munich, Germany
\email{choro.ulan-uulu@siemens.com} \and
Department of Computer Science and Engineering, Chalmers University of Technology, Gothenburg, Sweden \and
Department of Mathematics and Computer Science, Eindhoven University of Technology, Eindhoven, Netherlands
\email{j.bosch1@tue.nl} \and
Department of Computer Science and Media Technology, Malmö University, Malmö, Sweden
\email{helena.holmstrom.olsson@mau.se}}
\maketitle


\begin{abstract}

\gls{cae} enables simulation experts to optimize complex models, but faces challenges in user experience (UX) that limit efficiency and accessibility. While artificial intelligence (AI) has demonstrated potential to enhance CAE processes, research integrating these fields with a focus on UX remains fragmented.
This paper presents a multivocal literature review (MLR) examining how AI enhances UX in CAE software across both academic research and industry implementations. Our analysis reveals significant gaps between academic explorations and industry applications, with companies actively implementing LLMs, adaptive UIs, and recommender systems while academic research focuses primarily on technical capabilities without UX validation. 
Key findings demonstrate opportunities in AI-powered guidance, adaptive interfaces, and workflow automation that remain underexplored in current research. By mapping the intersection of these domains, this study provides a foundation for future work to address the identified research gaps and advance the integration of AI to improve CAE user experience.


\end{abstract}

\section{Introduction}

Computer-aided engineering (CAE) software employs mathematical models to predict system behavior across engineering domains \cite{GAROUSI2019101}, enabling virtual testing that reduces costs and accelerates development \cite{nvidiasimscale}. These tools have become essential in industries from aerospace to automotive, where engineers simulate everything from aerodynamics to manufacturing quality \cite{bios12070491}.

Despite their value, simulation tools present significant usability challenges. User experience (UX), defined by \cite{normalizacyjny_ergonomics_2011} as "A person's perceptions and responses resulting from the use and/or anticipated use of a product, system or service", is critical for the effective use of these tools. In the CAE context, effective UX enables engineers to perform complex simulation tasks—such as geometry preparation, mesh generation, physics setup, and results interpretation—with minimal friction, cognitive load, and potential for error. However, many CAE tools struggle in this regard: engineers must often create precise geometric models, accurately specify numerous parameters, and possess extensive domain knowledge—with incorrect settings potentially wasting hours of computation time \cite{GAROUSI2019101}. These UX challenges limit adoption and effective utilization of powerful simulation capabilities.
Artificial intelligence (AI) is transforming simulation workflows through several key interventions \cite{GAROUSI2019101}. Large language models like AnsysGPT \cite{ansys} provide around-the-clock technical guidance to engineers, while Siemens' Industrial Copilot \cite{indscop} enhances interface intuitiveness and reduces cognitive load. Such tools address critical UX barriers by making complex systems more accessible.
Beyond assistance, AI accelerates simulation through neural networks trained on previous results, enabling non-specialists to evaluate designs within minutes rather than hours. Ansys SimAI exemplifies this approach, allowing more design alternatives to be tested across development phases \cite{ansyssimai}. This democratization of simulation capability represents a major UX advancement.
Despite AI's demonstrated potential to alleviate CAE's significant UX challenges, a clear, synthesized understanding of how these advancements are currently being implemented and validated—both in academic research and industry practice—is lacking. It remains unclear whether academic explorations align with industry needs, which AI-driven UX enhancements are gaining traction, and what specific research gaps hinder the translation of potential into widespread, effective application. Addressing this knowledge gap is crucial for guiding future research and development efforts aimed at fully leveraging AI to improve CAE user experience.

This paper contributes by: (i) systematically analyzing AI advancements impacting CAE software UX; (ii) identifying gaps between academic research and industry implementation; and (iii) mapping underexplored areas requiring further investigation. To achieve these contributions, this study employs a Multivocal Literature Review (MLR) as its primary methodology. This approach is chosen as it allows for the integration of insights from diverse sources, specifically utilizing a component systematic literature review (SLR) of academic research alongside a component Grey literature review (GLR) of industry practices. Adopting the MLR framework is crucial because, while previous literature has examined CAE and AI integration \cite{GAROUSI2019101}, it often lacks rigorous methodology, comprehensive market analysis, or a specific focus on user experience factors. The MLR overcomes these limitations by systematically analyzing and synthesizing both academic and industry perspectives, providing robust and comprehensive insights into AI-enhanced CAE user experience within this rapidly evolving field.

\section{Methodology}
\label{slcrq}
This study employed a MLR methodology to investigate the integration of AI for enhancing UX in CAE software and drawing insights from both academic and industry sources. The MLR consists of two complementary components: a SLR examining scholarly research and a GLR analyzing market solutions and industry practices.

\subsubsection*{\textit{A. Research Question (RQ)}}
Bridging the gap between academic research and market solutions requires a well-structured research question that clearly defines the scope of investigation. To achieve this systematically, we applied the PICOC framework \cite{picoc} during question formulation. This framework helps define the key elements – Population, Intervention, Comparison, Outcome, and Context – ensuring clarity and focus, an approach suitable within software engineering contexts \cite{kitch}.
\newline
\newline
\textit{RQ:} How is AI implemented in CAE software to enhance user experience, and what scientific research approaches remain to be translated into market solutions?

\subsubsection*{B. Search strategy}
\label{searchstrategy}

\textit{1) SLR Search:} Based on its coverage and relevance for the topic of AI in CAE software for better UX, we chose \textit{Google Scholar} to find articles \cite{WOHLIN2020106366}.
The selected time frame spans from 2010 to March 13th, 2025. This selection ensures a manageable scope for the review process. The study applied the following search string:

\texttt{("CAE software" OR "engineering simulation") AND ("AI integration" OR "machine learning") AND ("UX" OR "user experience") AND ("industry case study" OR "implementation" OR "adoption challenges")}

\textit{2) GLR Search:} The GLR search strategy adhered to guidelines from \cite{GAROUSI2019101}. Key companies at the intersection of \gls{cae}, \gls{ai}, and UX were initially identified via Google. Their websites were subsequently reviewed for relevant literature. Finally, a broad keyword search was conducted until theoretical saturation was achieved \cite{GAROUSI2019101}.

\subsubsection*{\textit{C. Inclusion and Exclusion Criteria}}
This study formulated the inclusion and exclusion criteria for the SLR part of the MLR based on \cite{kitch}. Table \ref{tab:criteria_application} lists these criteria.

\begin{table}[h]
\centering
\caption{INCLUSION/EXCLUSION CRITERIA AND APPLICATION RESULTS}
\label{tab:criteria_application}
\textit{Note: Application of inclusion criteria resulted in 14 papers for quality assessment.}
\\[5pt]
\begin{tabular}{|p{1.4cm}|p{6.7cm}|p{1.8cm}|p{2.1cm}|}
\hline
\textbf{ID} & \textbf{Criterion} & \textbf{Absence} & \textbf{Exclusions} \\
\hline
\textit{IC1} & Publication date within defined time frame & \textit{EC1} & 0 \\
\hline
\textit{IC2} & Primary study (not secondary/tertiary) & \textit{EC2} & 38 \\
\hline
\textit{IC3} & Peer-reviewed publication \newline (not GL, blog posts, PhD Thesis, preprints) & \textit{EC3} & 57 \\
\hline
\textit{IC4} & Published in English & \textit{EC4} & 1 \\
\hline
\textit{IC5} & Not a duplicate of another included study & \textit{EC5} & 0 \\
\hline
\textit{IC6} & Explicitly discusses AI integration within CAE \newline software environments & \textit{EC6} & 117 \\
\hline
\textit{IC7} & Addresses UX in AI-integrated CAE software \newline OR implies enhanced UX from AI integration & \textit{EC7} & 1 \\
\hline
\end{tabular}

\end{table}

For GLR: \textit{(i) Authority of the producer}. Does the author have expertise in the area? (e.g., job title principal software engineer). \textit{(ii) Novelty}. Does it enrich or add something unique to the market? Does it strengthen or refute a current position?

\subsubsection*{D. Quality Assessment}
\label{qasec}

The 14 selected SLR papers underwent quality assessment (QA) using criteria from \cite{Kmet2004STANDARDQA} with a Likert scale \cite{Likert1932A}; papers scoring below 60\% were excluded. To evaluate the performance of an LLM on this specific QA task, a sequential assessment and comparison process was followed.

First, the QA criteria were provided to the LLM (Claude 3.7). The 14 papers were then uploaded, and the LLM was prompted to assess each paper according to these criteria, generating scores and text-based justifications. Subsequently, the researchers performed an independent manual assessment of the same 14 papers using the identical criteria. Finally, the LLM-generated scores and justifications were compared against the results of the manual assessment. The manual assessment served as the definitive benchmark for accuracy. Final inclusion decisions, based on the 60\% quality threshold, were made solely based on the manual assessment scores. Table \ref{tab:quality_assessment} summarizes the results from both methods; nine papers were included based on the manual evaluation.

\begin{table}[htbp]
    \centering
    \caption{Number of Studies per Quality Percentile Range}
    \label{tab:quality_assessment}
    \begin{tabular}{|p{3.5cm}|p{1.2cm}|p{1.2cm}|p{1.2cm}|p{1.2cm}|p{1.3cm}|}
        \hline
        \textbf{Assessment Method} & \textbf{0-60\%} & \textbf{60-70\%} & \textbf{70-80\%} & \textbf{80-90\%} & \textbf{90-100\%} \\
        \hline
        Claude 3.7 Assessment & 6 & 3 & 3 & 1 & 1 \\
        \hline
        Manual Assessment & 5 & 3 & 3 & 0 & 3 \\
        \hline
    \end{tabular}
\end{table}

This comparison revealed crucial inaccuracies in the LLM's assessment compared to the manual benchmark. Our evaluation indicates that while LLMs may aid screening, manual assessment remains essential for ensuring the rigor required in systematic reviews.

\subsubsection*{\textit{E. Data Extraction}} 

Publication details, research field, methods, and outcomes were extracted for the SLR following \cite{DYBA2008833}. The search after quality assessment resulted in 9 papers. The data from those papers was extracted manually. The GLR examination focused on: 1) materials from CAE software manufacturers, 2) identified AI applications enhancing UX in CAE, and 3) results obtained through broad Google searches.

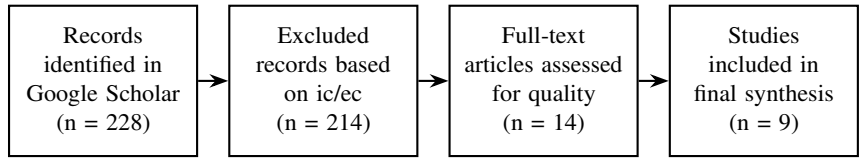
\begin{figure}
\centering
\begin{tikzpicture}[
    node distance = 0.4cm, 
    box/.style = {rectangle, draw, thick, minimum width=2.5cm, minimum height=2cm, 
                 text width=2.1cm, align=center, font=\footnotesize} 
]

\node[box] (box1) {Records identified in Google Scholar (n = 228)};
\node[box, right=of box1] (box2) {Excluded records based on ic/ec (n = 214)};
\node[box, right=of box2] (box3) {Full-text articles assessed for quality (n = 14)};
\node[box, right=of box3] (box4) {Studies included in final synthesis (n = 9)};

\draw[-{Stealth[length=2.5mm, width=1.8mm]}, thick] (box1.east) -- (box2.west);
\draw[-{Stealth[length=2.5mm, width=1.8mm]}, thick] (box2.east) -- (box3.west);
\draw[-{Stealth[length=2.5mm, width=1.8mm]}, thick] (box3.east) -- (box4.west);

\end{tikzpicture}
\caption{PRISMA Flow Diagram of SLR Selection Process}
\label{fig:prisma}
\end{figure}

\subsubsection*{\textit{F. Data Analysis and Synthesis}}

A qualitative synthesis, suitable for descriptive software reviews \cite{BRERETON2007571}, was applied. Reciprocal analysis \cite{Noblit1988MetaEthnographySQ} compared SLR and GLR findings to summarize gaps, complemented by descriptive synthesis.

\section{Results}
\label{sec:Results}
This section presents findings from the Multivocal Literature Review (MLR) on AI for enhancing CAE UX, synthesizing Systematic Literature Review (SLR) and Grey Literature Review (GLR) insights. A crucial SLR finding is the lack of published empirical UX evaluation for proposed AI methods, despite frequent claims of benefit. Industry practices, identified via GLR focused on market leaders (Siemens, Ansys, Altair, Dassault, PTC, Autodesk, Hexagon, Comsol) chosen for their influence, show a different emphasis: companies actively market AI features, explicitly articulating how specific capabilities (e.g., LLMs, automation) are designed to improve UX and providing the rationale. However, mirroring the academic gap in published proof, the reviewed grey literature generally lacks formal empirical validation results (like usability metrics) demonstrating these claimed outcomes, potentially due to competitive sensitivity. Growing academic publication volume since 2021 indicates increasing interest. Based on the combined MLR, key impact categories were identified; the following subsections detail SLR findings, GLR findings, and comparisons for each category, referencing summaries in Tables \ref{tab:ai_cae_ux}, \ref{tab:ai_cae_ux_grey}, and \ref{tab:comparison_grey_academic}."

\begin{table}[!t]
\centering
\caption{AI Methods in CAE for Enhanced User Experience: Systematic Literature Review}
\label{tab:ai_cae_ux}
\scriptsize
\renewcommand{\arraystretch}{1.05}
\begin{tabular}{|p{2.5cm}|p{2.4cm}|p{4.7cm}|p{0.7cm}|}
\hline
\textbf{Classification} & \textbf{AI technique} & \textbf{Paper Title (abbreviated)} & \textbf{Year}  \\
\hline
Workflow Automation & Conformal mapping & MeshLink: A surface structured mesh generation framework to facilitate automated data linkage & 2024  \\
\& Efficiency & Procedural Gen., ANN & Enhancing 3D Printing with Procedural Generation and STL Formatting Using Python & 2024  \\
\hline
Simulation Accel. & RL (mPPO, mHPPO) & RL approach with masked agents for chemical process design & 2024  \\
\& Optimization & PINN, DL & Physics-Informed DL for cyclone separators & 2021  \\
\hline
Generative Design \& Space Expl. & \multicolumn{3}{c|}{No direct examples in primary studies} \\
\hline
User Experience & \multicolumn{3}{c|}{No direct examples in primary studies} \\
\hline
Data-Driven Analysis & ML & Synthetic data generation for autonomous machinery & 2023  \\
\hline
System-Level Integration & KB (OWL), MAS & OWL ontologies for mixed-reality HRC assembly & 2023  \\
 & DT, Bayesian Opt. & Nominal digital twin for product design & 2023  \\
 & DL, GANs & Advanced HCI in digital twins & 2023  \\
\hline
Manufacturing \& QC & ML, DL, Image & ML-Enabled Prediction of 3D-Printed Microneedles & 2022  \\
\hline
Core AI Research & \multicolumn{3}{c|}{PINN, RL, DL Papers listed in categories above} \\
\hline
\end{tabular}
\end{table}

\textbf{Workflow automation and efficiency:} AI helps to automate tedious tasks that take time and could be automated. This generally leads to better user experience.

\textbf{SLR findings:}
\cite{ZHANG2024103661}  discusses the application of deep learning for generating surface structured meshes, automating a typically manual workflow step and potentially improving UX by reducing user effort. \cite{app14167299} demonstrates the use of procedural generation with AI to create CAD objects, suggesting enhanced UX through the generation of better models for 3D printing with less manual intervention.

\textbf{GLR findings:}
The GLR revealed five relevant industry examples in this area (Table \ref{tab:ai_cae_ux_grey}). Beta CAE utilizes AI in its RETOMO product for improved segmentation of CT images, enhancing UX by enabling analysis of noisy data \cite{betacaect}. They also offer an AI assistant for documentation navigation and automated CAE model generation from CAD \cite{betacaeaiassis}. Altair's ShapeAI employs ML for identifying matching parts \cite{altairshapeai}, PTC developed a Copilot for software development related to physical products \cite{ptc}, and Autodesk utilizes AI to interpret design markups and suggest context-aware actions \cite{autodescautocad}. These industry applications aim to augment user workflows, reduce redundant actions, and improve access to information, thereby enhancing overall user effectiveness.

\textbf{Comparison and gaps:}
A comparison between the GLR and SLR findings (summarized in Table \ref{tab:comparison_grey_academic}) indicates an overlap in automating preprocessing and design tasks. However, commercial efforts focus more broadly on integrating AI into the end-to-end workflow (e.g., CAD-to-CAE automation), whereas the reviewed academic studies concentrate on developing specific algorithms for tasks like mesh generation and procedural generation. A gap exists in academic research concerning the integration aspects and holistic workflow automation prevalent in industry applications.

\begin{table}[!t]
\centering
\caption{AI Methods in CAE for Enhanced User Experience: Grey Literature Review}
\label{tab:ai_cae_ux_grey}
\scriptsize
\begin{tabular}{|p{2.8cm}|p{2.1cm}|p{2.5cm}|p{3.1cm}|}
\hline
\textbf{Classification Group} & \textbf{Company} & \textbf{Product} & \textbf{Key AI techniques}  \\
\hline
Workflow Automation \&  & BETA CAE & Retomo & AI Segmentation \\
Efficiency & BETA CAE / Dassault & Ansa, Komvos & ML Toolkit  \\
 & ALTAIR & shapeAI & ML \\
 & PTC & Codebeamer Copilot & Generative AI \\
\hline
\makecell[tl]{Simulation\\ Acceleration \& }  & Ansys & Ansys Suite & Neural Networks \\
 Optimization & Ansys & Ansys SimAI & Physics-Informed AI \\
 & Neural Concept & NC Platform & 3D Deep learning / LLM  \\
 & ALTAIR & PhysicsAI & Geometric Deep Learning  \\
\hline
Generative Design \& Design Space Expl.  & Ansys & Ansys Discovery & HPC \& Cloud AI  \\
 & Autodesk & Autodesk Forma & Generative AI  \\
 & Autodesk & Autodesk Fusion & Generative AI \\
 & Neural Concept & NC & Deep Learning  \\
 & SimScale / Nvidia & PhysicsNeMo / SimScale & Physics AI / Eng. AI \\
 & Hexagon & Odysee CAE & AI/ML  \\
\hline
User Interaction \& Guidance & Ansys & AnsysGPT & Generative AI  \\
 & Ansys & Ansys SimAI & NLP  \\
 & Autodesk & Autodesk Maya & NLP  \\
 & BETA CAE & AI assistant & LLM  \\
 & SIEMENS & Simcenter Amesim & LLMs  \\
 & SIEMENS & NX & Adaptive UI (ML/AI)  \\
 & SIEMENS & Simcenter Amesim & Recommender LLM  \\
\hline
Data-Driven Analysis \&   & Ansys & Ansys SimAI & Machine Learning  \\
 Knowledge Management & Autodesk & Autodesk Fusion & Machine Learning  \\
 & Comsol & Comsol Multiphysics & LLMs / Supervised ML  \\
\hline
System-Level Integration
 & SIEMENS & Simcenter & Multi-Agent Systems  \\
 & Nvidia & Omniverse Blueprint & AI  \\
 & Dassault & 3D UNIV+RSES & Generative AI  \\
 \hline
Manufacturing \& Quality control 
& --- & --- & ---    \\
 \hline
Core AI Methodology Research & --- & --- & ---  \\
\hline
\end{tabular}
\end{table}

\begin{table*}[!htbp]
\centering
\caption{Comparison of Grey vs. Academic Literature by Classified Focus Area}
\label{tab:comparison_grey_academic} 
\footnotesize
\begin{tabular}{|p{2.0cm}|p{3.7cm}|p{3.0cm}|p{4.7cm}|} 
\hline
\textbf{Classification Group} & \textbf{GL Emphasis} & \textbf{Academic literature Emphasis} & \textbf{Overlaps and Gaps}  \\
\hline
Workflow Automation \& Efficiency & High focus: Automating CAD $\rightarrow$ CAE (BETA CAE), preprocessing (segmentation, part ID), integrating with related domains (software dev). & Present in preprocessing (MeshLink) and design automation (Procedural Generation). & \textbf{Overlap:} Automation of preprocessing (meshing/geometry prep) and design tasks. \textbf{Gap:} Commercial side addresses broader workflow integration (software dev, direct CAD $\rightarrow$ CAE automation). Academic focus on specific algorithms/frameworks. \\
\hline
\makecell[tl]{Simulation \\ Acceleration \& \\Optimization} & Significant focus: Using Physics-Informed AI, Deep Learning/LLMs, NN for faster simulations/predictions or enhanced accuracy. Key value proposition. & Significant focus: Developing/validating methods (PINNs, RL for optimization), demonstrating speed-up/accuracy on specific problems. Often methodological. & \textbf{Overlap:} Strong interest in Physics-Informed AI/PINNs, ML/DL for prediction/surrogates. \textbf{Gap:} Commercial side integrates into platforms; Academic side proves concepts/methods. \\
\hline
Generative Design \& Design Space Expl. & Strong focus: Using GenAI/AI/DL/ML for design creation/exploration, environmental analysis, optimization (Discovery, Forma, Fusion, NC, SimScale, Hexagon). Aided by ML toolkits (BETA CAE). & Less prominent in this specific academic sample, but exists in wider literature (e.g., topology optimization using AI). Often tied to optimization research. & \textbf{Overlap:} Use of AI for design exploration/optimization. \textbf{Gap:} Commercial side heavily pushing user-facing GenAI tools \& integrated toolkits for rapid exploration; Academic has developed algorithms for generating too but was excluded. \\
\hline
User Experience \& Guidance & High focus: NLP/LLM interfaces (support, interaction, search, code gen, replacing UI elements), Adaptive UI, Recommenders, NLP for scene interaction. Major area for UX improvement. & While literature exists on general CAE UX (excluding AI), the reviewed SLR papers showed minimal focus on specifically evaluating the UX impact of AI integration in CAE. & \textbf{Overlap:} Use of NLP/LLMs. \textbf{Gap:} Academic side has literature on LLMs for CAE but they do not mention UX. Commercial focus on LLMs/ML for diverse interaction tasks (support, search, UI replacement, coding, recommendations). \\
\hline
Data-Driven Analysis \& Knowledge Management & Present: Predictive modeling (performance, parameters), analysis tools (InfoDrainage), inferring properties (COMSOL viscosity example using ML/LLM). & Present: Prediction (microneedles), synthetic data generation, knowledge representation (Ontology in HRC paper). & \textbf{Overlap:} No overlap, diferent directions. \textbf{Gap:} Academic side explores foundational aspects like synthetic data and formal knowledge, software manufacturers focus on inferring viscosity from droplet shape. \\
\hline
System-Level Integration & Emerging focus: Digital Twins (Nvidia, Dassault), Agent systems (Siemens, COMSOL ASAs). Positioned as advanced/future capabilities. & Clear focus: Developing frameworks/concepts (NDT), enabling interactions (HRC ontology), specific system components (HAR in DT). & \textbf{Overlap:} Interest in Digital Twins and Agent concepts. \textbf{Gap:} Academic side often develops theoretical foundations for digital twins or models for human robot collaboration. \\
\hline
Manufacturing \& Quality Control & Less direct focus in the CAE software itself (which is pre-manufacturing), though outputs feed into it. & Present: Applying AI directly to manufacturing outcomes (microneedle quality). & \textbf{Gap:} This area is more prominent in manufacturing-specific research/tools than typical CAE UX focus, though linked via the design data. \\
\hline
\makecell[tl]{Core AI \\Methodology \\Research} & Not applicable (focus is on application). & Implicit in several papers (PINN development, RL agents, DL architectures for HAR). & N/A - This is primarily the domain of academic research. \\
\hline
\end{tabular}
\end{table*}

\textbf{Simulation acceleration and optimization:} AI optimizes parameters and simulation speed with physics, reducing user effort in software optimization.

\textbf{SLR findings:}
Two SLR papers addressed simulation acceleration and optimization. \cite{reynosodonzelli_reinforcement_2025} utilized masking techniques with RL agents to enhance the design and optimization of chemical process flowsheets, implying a UX benefit through reduced waiting times. \cite{QUEIROZ2021100002} demonstrated the use of physics-informed deep learning (PIDL) for predicting flow fields, claiming a significant UX impact through a model reported to be 200 times faster than traditional CFD simulations, thereby increasing productivity and user satisfaction.

\textbf{GLR findings:}
Industry sources indicate a strong focus on AI for simulation acceleration. Ansys developed SimAI, which facilitates AI-assisted user interaction and predicts design behaviors rapidly based on past simulations \cite{ansyssimai}. Ansys also published a case study on AI/ML integration for simulation optimization \cite{ansysaiml}. Neural Concept's NC Platform utilizes 3D deep learning and LLMs for accelerated design and simulation \cite{nc}. Altair's PhysicsAI employs geometric deep learning, enabling models to learn from geometry directly without extensive parametrization \cite{altairphysicsai}. These tools aim to empower users with faster results and higher quality insights, enhancing the overall user experience. 

\textbf{Comparison and gaps:}
A significant overlap exists between SLR and GLR in the application of AI (particularly PIDL/PINNs and deep learning) for prediction and surrogate modeling. However, a gap lies in the scope of prediction: SLR studies focused on specific phenomena (e.g., flow fields), whereas GLR examples target broader prediction of design behaviors. Furthermore, commercial entities focus on integrating these methods into comprehensive platforms, while academic studies often concentrate on methodological proof-of-concept.

\textbf{Generative Design and Design Space Exploration and User Experience and Guidance:} This combined category considers AI for automatically generating design variations (Generative Design), exploring the design space efficiently (Design Space Exploration), and enhancing user interaction through features like adaptive interfaces or intelligent guidance (User Experience and Guidance). 

\textbf{SLR Findings:}
The SLR yielded no primary studies specifically focused on Generative Design or AI-driven User Experience/Guidance within the defined scope. While papers generally alluded to improved user experience as an outcome of their proposed methods (e.g., through automation or speed), none provided empirical evaluations of UX or focused directly on AI for interaction design or guidance within CAE. This represents a notable gap in the reviewed academic literature.

\textbf{GLR Findings:}
In contrast, there is substantial industry activity in these areas which is shown through GLR findings. Autodesk's Fusion enables real-time 3D model creation and design exploration \cite{autodescautocad}. Neural Concept's NC platform assists users in product design, claiming AI as a key factor for successful products and improved UX \cite{ncdl}. SimScale and Nvidia collaborated on an AI Foundation Model for rapid exploration of design optimizations \cite{nvidiasimscale}. Hexagon developed Odysee for design space exploration and process optimization \cite{hexagon}, and Ansys offers Discovery for enhanced scalability and rapid design exploration \cite{ansysdisc}. For User Experience and Guidance, Ansys provides AnsysGPT for real-time simulation support \cite{ansys} and SimAI for interaction assistance. Beta CAE offers an AI assistant for documentation search and code generation \cite{betacaeaiassis}. Siemens is actively prototyping AI applications, including generative AI for UX enhancement across user groups \cite{kaiaiincae} and recommender systems \cite{kaiaiincaerecom}, aiming to reduce software complexity and improve accessibility.

\textbf{Comparison and Gaps:}
A clear disparity exists. A strong industry focus on using AI (especially Generative AI and LLMs) for design creation, exploration, and direct user guidance (support, adaptive UI, recommendations) is demonstrated in GLR results. The academic literature reviewed in the SLR lacks studies addressing the UX implications or evaluations of such systems within CAE environments. While academic research on generative algorithms exists, its integration and UX assessment within CAE tools appear underexplored in the SLR sample. Similarly, the use of LLMs/ML for direct user interaction tasks is prominent in GLR but largely absent from the specific focus of the reviewed SLR papers concerning UX impact.

\textbf{Data-Driven Analysis and Knowledge Management:}AI enables predictions and knowledge management, allowing early parameter adjustments and reducing simulation iterations.

\textbf{SLR findings:}
One SLR paper by \cite{SchusterHagmannsSonjiLöcklinPetereitEbertWeyrich+2023+953+968} highlighted that CAE data can be generated structured, enabling faster data acquisition for developers, which translates to higher productivity and reduced friction, implying a UX benefit.

\textbf{GLR findings:}
Industry examples include Ansys SimAI predicting design performance. Comsol Multiphysics applies ML for inferring material properties, such as viscosity from simulated droplet shapes, using inverse problem techniques \cite{comsol}.

\textbf{Comparison and gaps:}
When comparing the SLR and GLR results (Table \ref{tab:comparison_grey_academic}), its clear that the reviewed academic work emphasized aspects like synthetic data generation for model training and the potential for structured data acquisition. In contrast, the GLR examples focus on applying ML techniques within tools to infer specific properties or predict performance based on existing or simulated data. A gap appears in translating foundational academic work on data generation and formal knowledge representation into practical, user-facing prediction tools within commercial CAE software.

\textbf{System-Level Integration:} AI improves interaction with digital twins and enables multi-agent systems for better UX.

\textbf{SLR findings:}
The SLR included three papers on system-level integration. \cite{DAVID2023359} employed multi-agent systems for human-robot collaboration, suggesting improved UX through enhanced productivity. \cite{zhang_nominal_2023} demonstrated the feasibility and utility of specific digital twin types, potentially reducing user cognitive load by consolidating data. \cite{inbookahci} presented advanced Human-Computer Interaction (HCI) techniques, including human pose recognition, to improve interaction within digital twin environments, making systems feel more intuitive.

\textbf{GLR findings:}
Industry efforts focus on integrated platforms. Siemens is developing multi-agent integration within Simcenter Studio \cite{kaiaiincaema}. Nvidia's Omniverse Blueprint facilitates real-time CAE digital twins connected with commercial solvers \cite{nvidiaomni}. Dassault integrates generative AI into its virtual twin platform ("3D UNIV+RSES") \cite{dassault}. These initiatives aim to provide users with more comprehensive, interactive, and data-rich system representations.

\textbf{Comparison and gaps:}
Substantial literature exists in both SLR and GLR concerning AI agents and digital twins. A key difference lies in focus: the academic side often develops theoretical foundations or specific interaction models (e.g., for HRC or specific DT components). Industry concentrates on implementing these concepts within integrated, commercially-oriented platforms. The gap involves bridging foundational academic work with the practical implementation and UX considerations of interacting with these complex, integrated industrial systems.

\textbf{Manufacturing and Quality Control:} AI automates quality assessment of manufactured parts, streamlining the workflow.

\textbf{SLR findings:}
\cite{bios12070491} demonstrated the prediction of 3D-printed microneedle quality attributes (e.g., base diameter, height) using deep learning applied to design features. This represents a potential UX improvement by automating a quality control step.

\textbf{GLR findings:}
Autodesk documentation reflects the use of software for predicting manufacturability, integrating CAE with manufacturing considerations \cite{autodesccae}. However, specific GLR examples detailing AI integration within CAE for manufacturing quality control were less prominent compared to other categories.

\textbf{Comparison and gaps:}
While manufacturers acknowledge the link between CAE and manufacturing, the academic side provided a specific example of AI predicting quality from design data. The GLR suggests less direct integration of AI for QC within the core CAE UX workflow itself, possibly residing in downstream manufacturing tools. A potential gap exists in tightly coupling AI-driven manufacturability and QC predictions directly into the CAE interface for immediate user feedback during the design phase.

\textbf{Core AI Methodology Research:} Fundamental AI development using engineering problems as test cases to advance user experience.

\textbf{SLR findings:}
Several SLR papers, including those by \cite{QUEIROZ2021100002} on PIDL, \cite{reynosodonzelli_reinforcement_2025} on RL agents, and \cite{bios12070491} on ML for prediction, inherently contribute to this category by developing or applying advanced AI methods within a CAE context.

\textbf{GLR findings:}
The GLR does not typically reflect fundamental AI methodology research, as the focus of commercial software manufacturers is primarily on applying established or adapted AI techniques to deliver user value rather than publishing novel AI methods.

\textbf{Comparison and gaps:}
The comparison highlights the distinct roles: academia focuses on advancing AI methodologies, often validated using CAE problems, while industry concentrates on the application and integration of these methods. GLR emphasizes applicability and proven solutions, potentially hesitant to adopt untested novel AI approaches in the conservative CAE field. The primary "gap" is the inherent translation challenge in moving cutting-edge academic AI research into robust, scalable commercial software.

Based on this comprehensive review, considerable interest exists in leveraging AI to improve the user experience of CAE software. However, significant disparities remain between the focus of academic research and industry implementation. Notably, the evaluation of UX impacts through established metrics (e.g., usability evaluations, task completion times, adoption rates) appears limited in the reviewed academic literature. Several factors might explain this discrepancy: the rapid pace of software development, resource limitations hindering rigorous academic UX studies, and competitive sensitivity discouraging the publication of internal industry UX research.

\subsection{Mapping Underexplored Areas Requiring Further Investigation}
\begin{table*}[!htbp]

\centering
\caption{Potential AI Applications for Enhanced CAE UX, Informed by Methods Identified in Broader AI/UX Research \cite{uxai}}
\label{tab:stige}

\footnotesize

\begin{tabular}{|p{1.7cm}|p{2.19cm}|p{2.4cm}|p{4.1cm}|p{1.79cm}|}

\hline
\textbf{Classification Group} & \textbf{Potential AI method} & \textbf{Brief description} & \textbf{Potential relevance to CAE UX} & \textbf{Original phase} \\
\hline
Workflow Automation \& Efficiency & Sketch-to-Wireframe/Code Generation & Using ML/DNNs to convert hand-drawn sketches or low-fidelity mock-ups into digital wireframes or basic UI code. & Could potentially speed up early UI/workflow ideation for new CAE features or custom interfaces based on quick sketches from UX designers or even engineers. Its easier to draw a sketch and derive 3D CAD from it and turn it into CAE, then from plain text. & \makecell[tl]{Solution  \\ Design /\\ Development} \\

 & Code/GUI Generation from Visual Mockups & Using DL/Computer Vision to reverse engineer UI screenshots or high-fidelity mockups into functional code or GUI skeletons. & Although perhaps less common for complex CAE, could potentially automate parts of UI implementation if visual design mockups are created, bridging the design-development gap. & Development \\
\hline
Simulation Acceleration \& Optimization & Cognitive Modeling / Behavior Simulation & Using AI based on cognitive architectures (like ACT-R) to simulate human behavior and predict performance for different UI designs. & Could offer rapid, quantitative feedback on the usability of different CAE workflow designs or UI layouts by simulating how an engineer might interact with them, replacing some testing. & \makecell[tl]{Design \\ Evaluation} \\

\hline
\makecell[tl]{User \\Interaction \& \\Guidance} & \makecell[tl]{Design Pattern\\ Recommendation} & Using supervised ML to recommend appropriate interaction design patterns based on specified design-level requirements. & Could guide CAE developers/designers in selecting proven UI patterns for complex interactions (e.g., managing large parameter sets, visualizing multi-physics results, mesh setup). & \makecell[tl]{Solution \\ Design}  \\

 & \makecell[tl]{AI for Emotional\\ Design Aspects} & Considering AI's role within a design-oriented approach focused on emotional user experience. & Could guide the design of interfaces that reduce stress or enhance confidence for engineers using complex tools. & \makecell[tl]{Solution \\ Design} \\
\hline
Data-Driven Analysis \& Knowledge Management & Automated Persona Generation & Using ML and analytics data (e.g., online behavior) to automatically create user personas (AGP). & Could help CAE tool designers better understand different types of engineers using their software, potentially informing UI/feature design through data analysis. & Understanding Context of Use \\

 & Predicting Specific Usability Metrics & Using ML/DL to model and predict specific usability aspects like element tappability or interaction effort. & Could be adapted to analyze and predict usability challenges in dense CAE interfaces (e.g., discoverability of functions, perceived complexity) based on learned models. & \makecell[tl]{Design \\ Evaluation} \\
\hline
\end{tabular}
\end{table*}

This MLR has identified significant gaps between current industry applications and academic research concerning AI for enhancing UX in CAE. Specifically, the review highlighted areas where industry utilizes AI (e.g., LLM-based guidance, adaptive interfaces) that lack corresponding rigorous academic investigation or UX validation, alongside foundational academic work not yet fully translated into widespread practice (Section \ref{sec:Results} and Table \ref{tab:comparison_grey_academic}).

To explore how these identified gaps might be addressed and to map potential future directions, this section leverages findings from broader AI and UX research. Instead of conducting a separate, exhaustive literature search across the vast AI field, a highly relevant systematic literature review by Stige et al. \cite{uxai} was utilized as a proxy. This peer-reviewed SLR details how AI is leveraged across various stages of the general UX design process (based on 46 research articles).

While not focused specifically on CAE, the AI techniques and UX design principles examined in \cite{uxai} are often domain-agnostic. Core UX stages like "Understand the context of use," "Specify user requirements," and "Produce design solutions" are universally applicable. AI capabilities such as automation, recommendation, and content generation identified by Stige et al. can potentially be transferred to the specific context of CAE UX, even if implementation details differ. Therefore, \cite{uxai} provides a robust, systematically derived inventory of established and emerging AI methods relevant to enhancing UX, offering potential solutions for the gaps observed in the current CAE-focused AI literature (both SLR and GLR).

The following analysis involves extracting key AI applications identified as relevant to general UX design in the synthesis by Stige et al. \cite{uxai}. Our contribution then lies in analyzing the potential relevance and applicability of these broader AI methods specifically to the context of CAE UX, mapping them against the underexplored areas and needs identified in our MLR findings. This provides concrete opportunities for AI/UX experts and researchers to advance the field. The results of this mapping are presented in Table \ref{tab:stige}.

\section{Conclusion}

This work presented a systematic multivocal literature review (MLR) examining both academic research and grey literature (GL) on the use of Artificial Intelligence (AI) for enhancing User Experience (UX) in Computer-Aided Engineering (CAE) software.

Key findings reveal a significant disparity between academic focus and industry practice. Primarily, there is a notable lack of empirical UX evaluation in academic studies proposing AI methods for CAE, despite frequent author claims of UX improvement. Conversely, industry GL indicates active implementation of AI (especially Large Language Models (LLMs) and adaptive UI) aimed directly at enhancing UX, but this work often lacks formal publication detailing methodology or rigorous evaluation, potentially due to the competitive value of such research. Another finding is the limited academic focus on the specific UX implications of generative AI within the CAE manufacturing context. The comparison highlighted these divergences in focus and validation approaches between the two spheres.

The key contributions of this paper are threefold: (i) it provides the first comprehensive MLR specifically analyzing the intersection of AI, UX, and CAE from both academic and industry perspectives; (ii) it systematically identifies and categorizes current applications and research efforts, clearly articulating the critical gaps, particularly the deficit in empirical UX validation within academic publications; and (iii) it proposes concrete avenues for future research by leveraging broader AI/UX research \cite{uxai} to identify potential AI methods applicable to improving CAE UX (as detailed in Table \ref{tab:stige}).

By mapping these potential methods against the identified gaps, this work suggests pathways to advance user-centric CAE tool development and bridge the gap between academic potential and industry needs.

\subsection{Limitations and Future work}
The search strategy for this review intentionally focused on "CAE AI UX" and publications between 2010 and March 2025 for focused analysis within practical constraints. We acknowledge significant contributions from related areas like "interaction design" and "interface design," but a dedicated search incorporating these broader terms was beyond the current study's scope. Building upon the understanding gained from this review, future research will transition from analysis to creation, developing novel methods and potentially prototype solutions aimed at tangibly improving the user experience of AI capabilities within CAE software. This work will directly address the gaps and leverage opportunities identified herein.

\section{Acknowledgments}
The authors gratefully acknowledge the support provided by the Software Center. The authors also thank their colleagues and supervisors within Siemens
for valuable discussions and feedback during this research.

\bibliographystyle{abbrv-doi}

\bibliography{template}
\end{document}